\RequirePackage{ifpdf}
\RequirePackage{color}
\documentclass[12pt,letterpaper]{JHEP3}         
\usepackage{amssymb,amsfonts}

\usepackage{wrapfig}
\usepackage{epsfig}

\def\be{\begin{eqnarray}}
\def\ee{\end{eqnarray}}
\newcommand{\nn}{\nonumber}
\newcommand\para{\paragraph{}}

\newcommand{\eqn}[1]{(\ref{#1})}

\def\Dslash{\,\,{\raise.15ex\hbox{/}\mkern-12mu D}}
\def\Dbarslash{\,\,{\raise.15ex\hbox{/}\mkern-12mu {\bar D}}}
\def\delslash{\,\,{\raise.15ex\hbox{/}\mkern-9mu \partial}}
\def\delbarslash{\,\,{\raise.15ex\hbox{/}\mkern-9mu {\bar\partial}}}
\def\pslash{\,\,{\raise.15ex\hbox{/}\mkern-9mu p}}
\def\calDslash{\,\,{\raise.15ex\hbox{/}\mkern-12mu {\cal D}}}

\newcommand{\rmd}{{\rm d}}

\def\lae{\mathrel{\mathop{\smash{\lower .5 ex \hbox{$\stackrel<\sim$}}}}}
\def\lae{\mathrel{\mathop{\smash{\lower .5 ex \hbox{$\stackrel>\sim$}}}}}

\title{\mbox{Mirror Symmetry and Bosonization in 2d and 3d}}

\author{Andreas Karch${}^1$, David Tong${}^2$ and Carl Turner${}^2$\\
${}^1$Department of Physics, \\
University of Washington, Seattle, WA 98195, USA \\
${}^2$Department of Applied Mathematics and Theoretical Physics, \\
University of Cambridge, Cambridge, CB3 OWA, UK
}

\abstract{We study a supersymmetry breaking deformation of the ${\cal N}=(2,2)$ cigar = Liouville  mirror pair,   first introduced by Hori and Kapustin. We show that mirror symmetry flows in the infra-red to 2d bosonization, with the theories reducing to  massive Thirring and Sine-Gordon respectively. The exact bosonization map emerges at one-loop. We further compactify non-supersymmetric 3d bosonization dualities on a circle and argue that these too flow to 2d bosonization at long distances.}

\begin{document}
\pagestyle{plain} \setcounter{page}{1}
\newcounter{bean}
\baselineskip16pt \setcounter{section}{0}

\section{Introduction}

The purpose of this paper is twofold. First, we study the fate of 2d mirror symmetry subject to supersymmetry breaking deformations, and show that it reduces to 2d bosonization. Second, we study the fate of 3d bosonization upon compactification on a circle and show that this too reduces to 2d bosonization. 

\para
In Section \ref{hksec}, we focus  on a mirror pair first introduced by Hori and Kapustin \cite{hk}. This is a duality between  2d ${\cal N}=(2,2)$ supersymmetric theories which, colloquially, is written as
\be \mbox{Cigar}\ \ \ \longleftrightarrow\ \ \ \mbox{Liouville}\nn\ee
 We  systematically break supersymmetry on both sides of this duality. We will see that the two theories flow to the well known 2d bosonization duality \cite{coleman}, 
 \be \mbox{Massive Thirring}\ \ \ \longleftrightarrow\ \ \ \mbox{Sine-Gordon}\nn\ee
We show that the map between parameters in the bosonization duality descends from the mirror map after a one-loop shift.

\para
In Section \ref{3d2dsec}, we change tack somewhat and study a non-supersymmetric, 3d bosonization duality. This is not quite so divorced from the previous topic as it may seem. Both Hori-Kapustin duality and 3d bosonization can be derived from a common progenitor: a 3d ${\cal N}=2$ Chern-Simons matter mirror pair, first postulated in \cite{dt}. It has long been known  that, when compactified on a circle, 3d mirror symmetry gives rise to Hori-Kapustin duality \cite{ahkt}.  More recently it was shown that, when supersymmetry is broken 3d mirror symmetry gives rise to 3d bosonization \cite{shamit1,shamit2}. Our goal in the second half of the paper is to complete the circle of ideas. We argue that, upon compactification, 3d bosonization reduces to the better studied 2d bosonization.

%
 %
 %
%


\section{From 2d Mirror Symmetry to 2d Bosonization}\label{hksec}
\def\Imag{\mathrm{Im}\, }
\def\Real{\mathrm{Re}\, }

In this section, we treat each side of the Hori-Kapustin duality in turn. Our strategy  for breaking supersymmetry follows \cite{shamit1,shamit2}, where the fate of 3d mirror pairs was studied (a story we will return to in Section \ref{3d2dsec}).  We identify a global symmetry on each side of the duality and couple this to a background vector multiplet. Turning on scalar and   $D$-term parameters in this vector multiplet allows us to map supersymmetry breaking deformations across the duality.  We will show that the cigar theory flows to the massive Thirring model, while the Liouville theory becomes Sine-Gordon.

\para
Before we proceed, we pause to make a comment. There is a natural non-supersym- metric analog  of the Hori-Kapustin mirror pair: FZZ duality \cite{fzz}. This is an equivalence between the bosonic cigar and sine-Liouville theory, proven in \cite{schomerus}. One might have thought that breaking supersymmetry would reduce Hori-Kapustin duality to FZZ. However, this turns out not to be the case. This is similar in spirit to what happens in three dimensions where one might have thought that breaking supersymmetry would reduce 3d mirror symmetry to its non-supersymmetric counterpart, particle-vortex duality. Yet, this is not how things pan out in either case. Instead, in both 2d and 3d, breaking supersymmetry reduces mirror symmetry to  bosonization.

\subsubsection*{The Cigar}

The cigar theory is a non-linear sigma model whose target space has the shape of a semi-infinite cigar. It can also be viewed as an $SL(2,{\bf R})/U(1)$ coset superconformal field theory. The 2d fields consist of a complex scalar $\phi$, and a single Dirac fermion $\psi$, with Lagrangian\footnote{The change of coordinates $\rho = \sinh|\phi|$ and $\alpha = {\rm arg}\phi$ results in the more familiar cigar metric $ds^2 = d\rho^2 + \tanh^2\!\rho\, d\alpha^2$.}
\be
{\cal L}_{\rm cigar} = \frac{\gamma}{4\pi}\left[ \frac{1}{1+|\phi|^2} \,|\partial_i\phi|^2 + \frac{1}{1+|\phi|^2}\, i\bar{\psi}\!\Dslash \psi + \frac{1}{4(1+|\phi|^2)^3}\psi\psi \bar{\psi}\bar{\psi}\right]
\label{cigar} \ee
The theory has a marginal coupling,  $\gamma$. In the asymptotic region, $|\phi|\rightarrow \infty$, the target space is a cylinder with radius  $\sqrt{\gamma/4\pi}$. As usual in a supersymmetric sigma-model, the coefficient of the four-fermi interaction is governed by the Riemann tensor of the target space.  The theory is weakly coupled when 
\be \gamma \gg  1\nn\ee
Geometrically, this arises because the Ricci curvature at the tip of the cigar is proportional to $1/\gamma$. We study the cigar theory only in this perturbative regime. 
 
 \para
The cigar target space is not Ricci flat so does not, on its own, give rise to a CFT. One compensates for this by including a background dilaton, although we  will not need this for our purposes. The supersymmetric cigar CFT plays a number of interesting cameo roles in string theory. One particularly interesting feature arises in the elliptic genus, which suffers a holomorphic anomaly and is related to the study of mock modular forms \cite{jan,eguchi,sujay,sameer}.

\para
The cigar theory has a $U(1)_V$ global symmetry, under which both $\phi$ and $\psi$ have charge $+1$. There is also a $U(1)_R$ R-symmetry, under which $\psi$ transforms while $\phi$ does not.  For the purposes of supersymmetry breaking,  our interest lies in the $U(1)_V$ global symmetry, which we couple to a background vector superfield $V$. In  two dimensions, this can alternatively be packaged as a twisted chiral multiplet of the form
\be \Sigma = \frac{1}{\sqrt{2}}\bar{D}_+D_-V = m + \ldots + \vartheta^+\bar{\vartheta}{}^-(D-iF_{01})\nn\ee
(Note that we refer to superspace coordinates as $\vartheta$ to avoid confusion with the Sine-Gordon scalar $\theta$ that will make an appearance later.)
The background gauge field $A_i$, with field strength $F_{01}$, couples to the $U(1)_V$ current; we will have need for this when we come to compare the symmetries across the duality. In the meantime, our interest lies in the coupling of the parameters $m\in {\bf C}$ and $D\in {\bf R}$. Turning these on deforms the theory by 
\be \Delta {\cal L}_{\rm cigar} =  D|\phi|^2  +  \frac{|m|^2|\phi|^2}{1+|\phi|^2} +  \frac{m}{1+|\phi|^2}\bar{\psi}\psi\nn\ee
Note that supersymmetry is preserved when $m\neq 0$, and the theory becomes gapped, with the ground state at the tip of the cigar, $\phi=0$. In contrast, when $D\neq 0$ there is again a unique  ground state at $\phi=0$, but now supersymmetry is broken.

\para
In what follows, we take $D\gg m^2 \in {\bf R}^+$. (The more general case of $m\in {\bf C}$ follows from axial rotations of the fermions.) The scalars can now be integrated out. At tree level, we are left with the massive fermions interacting through a four-fermion term. In 2d, there is a unique four-fermion coupling and it can equivalently be written in Thirring form, as a current-current interaction. After canonically normalising the kinetic term, 
the cigar Lagrangian \eqn{cigar} becomes
\be
{\cal L}_{\rm Thirring} =  i\bar{\psi}\!\delslash\psi  + m  \bar{\psi} \psi  - \frac{\pi}{\gamma}
(\bar{\psi}\gamma^\mu\psi)(\bar{\psi}\gamma_\mu\psi) \label{thirring}
\ee
When $m=0$, this describes a $c=1$ conformal field theory, with $\gamma$ a marginal parameter. 

\subsubsection*{Liouville Theory}

The mirror of the cigar is Liouville theory \cite{hk}. It involves a single, twisted chiral superfield $Y$ with lowest component
\be Y = (y + i\theta) + \ldots\ \ \ \ {\rm with}\ \theta\in [0,2\pi)\nn\ee
Note that the compact scalar ${\rm Im}\, Y =\theta$ should not be confused with the superspace coordinate  $\vartheta$. The Lagrangian is
\be
{\cal L} _{\rm Liouville}= \int \rmd^4\vartheta \  \frac{1}{ 4\pi\gamma}  |Y|^2 +    \frac{\mu}{4\pi}\left[ \ \int \rmd^2\tilde{\vartheta} \  e^{-Y} + \mbox{h.c.} \ \right] \nn \ee
Here $\mu$ is an ultra-violet scale. (For example, in the Hori-Kapustin derivation of the duality, the superpotential arises from integrating out vortices which have a UV structure at the scale $\mu$.)

\para
The $1/4\pi \gamma$ normalisation of the kinetic term is important. The fact that this is the correct normalisation can be seen by examining the $y\rightarrow \infty$ limit of configuration space, where the theory is again described by a cylinder, now with radius $1/\sqrt{4\pi \gamma}$. This is the T-dual of the asymptotic cigar. 

\para
In contrast to the cigar, the geometry does not disappear when $y<0$; instead this regime is energetically disfavoured by the superpotential term.  The Liouville theory is under perturbative control when
\be \gamma \ll 1\nn\ee
This is the opposite regime from the cigar, as appropriate for a strong-weak duality.

\para
The Liouville theory also enjoys both a $U(1)_V$ and $U(1)_R$ symmetry. The R-symmetry involves a phase rotation of the fermion, coupled with a shift of the scalar $\theta$, so that the Yukawa coupling $e^{-y-i\theta}\psi\psi$ remains invariant. Here our interest lies in the $U(1)_V$ global symmetry which now manifests itself as a topological winding symmetry, with bosonic current
\be j^i = \frac{1}{2\pi}\epsilon^{ij}\partial_j\theta\label{jtop}\ee
As in the cigar theory, we couple this current to a background superfield $V$. This is achieved by adding the deformation
%
%
%
\be \Delta {\cal L}_{\rm Liouville} =  \frac{1}{2\pi}
\int \rmd^4 \theta \ VY + \mbox{h.c.}  = \frac{1}{4\pi}\int \rmd^2 \tilde
{\vartheta} \
\Sigma Y + \mbox{h.c.}
\nn \ee
The expansion includes the theta term $\theta F_{01}$, where $\theta$ is dynamical and $F_{01}$ is the background field strength. After an integration by parts, this gives the required $A_i j^i$ coupling, where $j^i$ is the topological current \eqn{jtop}.

\para
It is simple to trace the effect of the background $m$ and $D$ parameters in $\Sigma$. The twisted mass $m$ results in a  linear shift to the twisted superpotential $\Delta W = mY$. Including both parameters, we are left with the scalar potential
\be
V(Y) = \left|- \frac{\mu}{2} e^{-y-i\theta} + m \right|^2 + D y
\nn\ee
We again work in the regime $\mu^2 \gg D\gg m^2 \in {\bf R}^+$. The minimum sits at
\be e^{-y} \approx \frac{\sqrt{2D}}{\mu}\label{yground}\ee
In this ground state, fluctuations of both the real field $y$ and the fermion get a mass of order $\sqrt{D}$. We integrate them out. Meanwhile ${\rm Im}\,Y =\theta$ remains light.  At tree-level, the low-energy dynamics of theory is governed by the Sine-Gordon model, with
\be
{\cal L}_{0} = \frac{1}{4\pi \gamma} (\partial_i \theta)^2 + \sqrt{2 D} |m| \cos
\theta\label{sgtree}\ee
Here, the susy breaking scale $D$ should be thought of as the UV cut-off of the theory.

\para
The effective action \eqn{sgtree} arises at tree-level. It is a simple matter to compute one-loop corrections. Those of interest arise from the Yukawa coupling $e^{-y-i\theta}\psi\psi$. In the ground state \eqn{yground}, with canonically normalised kinetic terms,  this takes the form
\def\fmass{M}
\be {\cal L}_{\rm Yukawa} = i \bar{\psi}\!\delslash\psi + \fmass \left(e^{-i\theta} \psi \psi + \rm{h.c.} \right) \nn \ee
where $\fmass \sim {\cal O}(\sqrt{D})$ is a large mass scale. The exact value of $M$ will not matter for our purposes but, as we will see, the vertex gives a finite correction to the propagator for the light, periodic scalar $\theta$.  To proceed, we decompose the 
Dirac spinor $\psi$ into Majorana-Weyl components $\chi$, and write
\be {\cal L}_{\rm Yukawa} = i\chi^T\Delta_\theta\chi\nn\ee
with
\def\cosT{M\cos\theta}
\def\sinT{M\sin\theta}
\be \psi= \left(\begin{array}{c} \chi_1 + i\chi_2\nn\\ \chi_3 + i\chi_4\end{array}\right)\ \ {\rm and}\ \ \
\Delta_{\theta} = 
\left(
\begin{array}{cccc}
								&  i(\partial_0 - \partial_1) 	& -\sinT      					& \cosT \\
  -i(\partial_0 - \partial_1)	&       						& \cosT							& \sinT   \\
    \sinT						& -\cosT    					& 								& i(\partial_0 + \partial_1)   \\
    -\cosT						& -\sinT    					& -i(\partial_0 + \partial_1) 	&   \\
\end{array}
\right)
\nn\ee
Expanding about $\theta=0$, the contribution to the low-energy, effective Euclidean action is
\be \delta S^{(E)} &=& - \left[ \log \mathrm{Pf}\, \Delta_{\theta} -  \log \mathrm{Pf}\, \Delta_{0} \right] \nn \\
&=& -\frac{1}{2} {\rm Tr} \left[ \Delta_{0}^{-1} (\Delta_{\theta} - \Delta_{0}) - \frac{1}{2}\Delta_{0}^{-1} (\Delta_{\theta} - \Delta_{0})\Delta_{0}^{-1} (\Delta_{\theta} - \Delta_{0}) + \cdots \right]
\nn \ee
The terms of interest are quadratic in the Fourier mode $\theta(p)$, and come from the diagram
\be \raisebox{-1.6ex}{\epsfxsize=1.2in\epsfbox{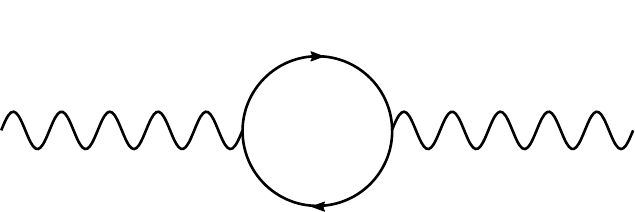}}\ =\  \frac{1}{4} \int \frac{\rmd ^2 p}{(2\pi)^2} \tilde{\theta}(p)\tilde{\theta}({-p})  \int \frac{\rmd ^2 q}{(2\pi)^2} \frac{-2\fmass^2(2\fmass^2 + 2(p+q)\cdot q)}{(q^2+\fmass^2)((p+q)^2+\fmass^2)}  \nn \ee
Naively, this contributes a mass term $\sim {\cal O}(M)$ for the scalar $\theta$. This, however, is an artefact of the expansion around $\theta=0$. Indeed, the R-symmetry, which acts as an axial rotation of $\psi$, together with a shift of $\theta$, prohibits the generation of a potential for $\theta$ in the limit $m=0$. In contrast, we are interested in the contribution to the kinetic terms. These arise by isolating the order $p^2$ term, and are given by
\be \delta S^{(E)}_{\rm kin}
&=& \frac{1}{4} \int \frac{\rmd ^2 p}{(2\pi)^2}\ \tilde{\theta}(p)\tilde{\theta}(-p)
\int \frac{\rmd ^2 q}{(2\pi)^2}\ \frac{-2\fmass^2(-2\fmass^2 p^2 + 4(p\cdot q)^2 - 2p^2q^2)}{(q^2+\fmass^2)^3} \nn \\
&=& \frac{1}{4} \int \frac{\rmd ^2 p}{(2\pi)^2}\ p^2 \tilde{\theta}(p)\tilde{\theta}(-p) 
\int \frac{\rmd ^2 q}{(2\pi)^2}\ \frac{4\fmass^4}{(q^2+\fmass^2)^3} \nn \\
&=& \frac{1}{8\pi} \int \rmd^2 x \ (\partial \theta)^2
\nn\ee
Combining this with the tree level result \eqn{sgtree}, the one-loop effective action is the Sine-Gordon model, with action
\be
{\cal L}_{SG} = \left(\frac{1}{4\pi \gamma} + \frac{1}{8\pi}\right)\,(\partial_i \theta)^2 + \sqrt{2 D} |m| \cos
\theta\label{sg}\ee

\para


\subsubsection*{Comparison to Bosonization}

We now compare these results with the 2d bosonization map of Coleman \cite{coleman}. He showed that the massive Thirring model, with Lagrangian 
\be {\cal L} =  \ i \bar{\psi}\!\delslash\psi + m\bar{\psi}\psi - g (\bar{\psi}\gamma^\mu\psi)(\bar{\psi}\gamma_\mu\psi) \nn\ee
is equivalent to the Sine-Gordon model, with Lagrangian
\be {\cal L} =   \frac{\beta^2}{2} (\partial_i\theta)^2 + \Lambda m \cos \theta \nn\ee
where $\Lambda$ is an ultra-violet scale, and the coupling constants in the two theories are related by
\be \beta^2 = \frac{1}{4\pi} + \frac{g}{2\pi^2} \label{good}\ee
Our supersymmetric mirrors had a marginal coupling $\gamma$. The cigar theory is weakly coupled when $\gamma\gg 1$ and flows to the Thirring model \eqn{thirring}, with coupling
\be \frac{g}{2\pi^2} = \frac{1}{2\pi \gamma} + O\left(\frac{1}{\gamma^2}\right) \nn \ee
where the correction comes from possible one-loop effects or higher. Meanwhile, the Liouville theory is weakly coupled when $\gamma \ll 1$ and flows to the Sine-Gordon model \eqn{sg}, with coupling
\be \beta^2 = \frac{1}{2\pi\gamma} + \frac{1}{4\pi} + {\cal O}(\gamma)\label{sgshift}\ee
where now the correction comes from possible two-loop effects or higher. 

\para
We see that the one-loop shift of the Sine-Gordon coupling \eqn{sgshift} reproduces exactly Coleman's bosonization map \eqn{good}.  This suggests that the higher-loop contributions to the map vanish. Indeed, other  aspects of the Sine-Gordon model have long been known to be  one-loop exact \cite{dhn}.

\section{From 3d Bosonization to 2d Bosonization}\label{3d2dsec}

There is a derivation of Hori-Kapustin duality that has its origin in three dimensional physics.\footnote{In fact, there are (at least) two such derivations. In \cite{mirrorwall}, it was shown the ${\cal N}=(2,2)$ theories naturally live on domain walls in the self-mirror 3d ${\cal N}=4$ theory. Viewed from the perspective of the Higgs branch variables, the dynamics of the domain walls is described by the cigar theory; viewed from  the Coulomb branch, one finds the mirror Liouville description.} One can start from the class  ${\cal N}=2$, 3d Chern-Simons matter mirror pairs, first  introduced in \cite{dt,me}. The simplest of these mirrors is:
\be \mbox{Free Chiral} \ \ \ \longleftrightarrow\ \ \  U(1)_{1/2} + \mbox{chiral} \label{3ddual}\ee
The first hint that this is related to Hori-Kapustin mirror pairs comes from examining the Coulomb branch of the right-hand side which, at finite coupling, has the shape of a cigar. 

\para
In \cite{ahkt}, 3d ${\cal N}=2$ mirror pairs were compactified on a spatial circle of radius $R$. (See also \cite{ofer3d2d} for a more recent discussion.)  In the present context, one achieves this by first extending the duality to the ``all-scale mirror symmetry" of \cite{ks}, which means replacing the left-hand side of the duality above with the gauged linear model whose Higgs branch is the cigar \cite{hk}. One finds that the 3d mirror pair \eqn{3ddual} reduces to the Hori-Kapustin duality in two dimensions. The marginal parameter $\gamma$ in the 2d theories is identified as $\gamma = e^2R$, where $e^2$ is the 3d gauge coupling.

\para
Alternatively, one may also study the fate of the mirror pair \eqn{3ddual} upon breaking supersymmetry. This was done in \cite{shamit1,shamit2}. (A related discussion for large $N$ bosonization dualities was given in \cite{jain,guy3}.) The result, assuming a lack of phase transitions under RG, is that the 3d mirror pair flows to a non-supersymmetric  3d bosonization duality, which takes the form
\be  \mbox{free Dirac fermion}   \ \ \  \longleftrightarrow \ \ \ \mbox{$U(1)_{1}$ coupled to XY critical point}\label{theduality}\ee
This duality has been the subject of much interest.  The fact that one can employ Chern-Simons terms to transmute the statistics of particles has long been appreciated in the context of non-relativistic physics or, relatedly, in quantum field theories with a gap \cite{wzee,polyakov,fs}. Evidence for such a duality at the critical point has come only recently, starting with large $N$ bosonization dualities inspired by holography  \cite{shiraz,ofer1}, for which  overwhelming evidence has been amassed \cite{guy,ofermore,shiraz2,shiraz3,ofer,shiraznew}. For the simple, Abelian bosonization duality \eqn{theduality}, there are now both lattice \cite{lattice} and coupled wire \cite{thewire} constructions. Moreover, it can be shown  \cite{kt,ssww} that the relationship \eqn{theduality} implies a host of other dualities, among them the long-established particle-vortex duality \cite{peskin,dh}. 

\para
In this section, we  study the fate of 3d bosonization  when both sides of \eqn{theduality} are compactified on a spatial circle ${\bf S}^1$ of radius $R$. (The fate of bosonization when placed on a manifold with boundary was considered in \cite{andreas1,andreas2}.)
At low energies, each side reduces to a theory in $d=1+1$ dimensions. 
We would like to identify these theories.

\para
The fermionic theory is, of course, trivial.  We introduce a mass term, $m_F\ll 1/R$, for the 3d fermion. We give the fermions periodic boundary conditions on the ${\bf S}^1$, leading to a tower of Kaluza-Klein (KK) modes with masses $|n/R+im_F|$ with $n\in {\bf Z}$. At low energies, only the zero mode survives, leaving us with a single two-dimensional Dirac fermion,
\be S_{\rm 2d} =   \int \rmd^2x\ i \bar{\psi}\!\delslash\psi + m_F\bar{\psi}\psi \label{freefermion}\ee
This  is well known to be dual to a compact scalar (for a special value of the radius).  Our task is to understand how this might arise from the 3d bosonic theory.

\para
This is not a simple question to address. The 3d bosonic theory is strongly coupled, and we have little control over its dynamics. Indeed, if we had any control, it would presumably come as less of a surprise to learn that the theory is actually a free Dirac fermion in disguise.

\subsection*{A Naive Approach: Compactification at Weak Coupling}

We start by describing an approach that does not, ultimately, work. We work with the 3d $U(1)$ gauge theory,
\be S_{\rm 3d} = \int \rmd^3x \ -\frac{1}{4e^2} f_{\mu\nu}f^{\mu\nu} + |{\cal D}_\mu\phi|^2 -m^2_B |\phi|^2 - \lambda|\phi|^4 + \frac{1}{4\pi}\epsilon^{\mu\nu\rho}a_\mu\partial_\nu a_\rho\nn\ee
This  theory is weakly coupled in the UV, but  flows to the strongly coupled IR fixed point that we wish to study. It has two, dimensionful couplings, namely $e^2$ and $\lambda$. When compactified on a circle, we  can construct two dimensionless couplings,
\be \gamma = e^2R\ \ \ {\rm and}\ \ \ \hat{\lambda} = \lambda R\nn\ee
We would like to understand the theory in the regime $\gamma \gg 1$ and $\hat{\lambda} \sim {\cal O}(1)$, where it first flows to the strongly coupled 3d fixed point, and is subsequently compactified on ${\bf S}^1$. Instead, we have control  in the regime $\gamma \ll 1$ and $\hat{\lambda}\ll 1$ where the theory is weakly coupled and we can do perturbative calculations. One may hope that there is no phase transition as we vary $\gamma$ and $\hat{\lambda}$, so that we can still see evidence of the bosonization duality in this regime. As we now show, this hope is ill-founded.

\para
Upon compactification, the 3d gauge
field decomposes into a 2d gauge field $a_i$, $i=0,1$  and the Wilson line
\be \theta = \oint a\ \in [0,2\pi)\label{wline}\ee
The Maxwell and Chern-Simons terms become
\be \int \rmd^3x\ -\frac{1}{4e^2}f_{\mu\nu}f^{\mu\nu} +
\frac{1}{4\pi}\epsilon^{\mu\nu\rho}a_\mu \partial_\nu a_\rho = 
\int \rmd^2x\   \frac{1}{4\pi \gamma} (\partial_i\theta)^2 + \frac{1}{2e^2_{\rm 2d}}
f_{01}^2 + \frac{1}{2\pi}\theta f_{01}\nn\ee
with $e^2_{\rm 2d} =  e^2/2\pi R$. We see that the Wilson line acts as a dynamical theta term in the 2d theory.  Meanwhile, the scalar field $\phi$ decomposes into an infinite tower of KK modes, together with the surviving zero mode of mass $m_B\ll 1/R$. When $\gamma,\hat{\lambda}\ll 1$, we may integrate out the tower of KK modes. A standard one-loop calculation (see, for example, \cite{gpy}) shows that these generate a potential for the Wilson line. When $m_B=0$, this potential is
\be V_{\rm eff} \sim  \frac{1}{2\pi R^2} \sum_{n=1}^\infty \frac{\cos(\theta n)}{n^3}\label{kkpot}\ee
 (A similar result holds when $m_B \ll 1/R$.) This gives the Wilson line a mass of order $1/R$, freezing it to its minimum which sits at $\theta=\pi$. The upshot is that, at low energies,  we're left with the Abelian-Higgs model in $d=1+1$ dimensions, with $\theta=\pi$. 

\para
The Abelian-Higgs model in $d=1+1$ dimensions has an interesting phase structure at $\theta=\pi$.   For $m_B^2\gg e^2_{\rm 2d}$ we can integrate out the scalar field and the resulting Maxwell theory has two degenerate ground states, corresponding to the electric field $E = \pm e_{\rm 2d}^2/2$. Meanwhile, when $m_B^2\ll - e^2_{\rm 2d}$, the theory lies in the Higgs phase and there is a unique ground state. This means that as we vary $m_B^2$ from positive to negative, we pass through a critical point. At this critical point, the theory is described by the Ising CFT with central charge $c=1/2$ \cite{wittencpn,affleck,nahum,zoharcpn}. 

\para
Interesting as this phase structure is, it does not coincide with the free Dirac fermion \eqn{freefermion} where, as $m_F$ is varied, we pass through a $c=1$ critical point. Clearly the weak coupling physics, when $\gamma,\hat{\lambda}\ll 1$, is rather different from the strong coupling physics of the duality. 

\subsection*{Compactification at Strong Coupling}

To proceed, we need a better handle on the strongly coupled fixed point of the bosonic theory. Fortunately we have one: the duality! Our goal, after all, is to use knowledge of the 3d duality to say something about the theories after compactification. 

\para
In fact, much of what we need follows immediately from the map between current operators in the 3d theories,
\be J_{\rm 3d}^\mu = \bar{\psi}_{\rm 3d}\gamma^\mu \psi_{\rm 3d}\ \ \longleftrightarrow\ \ J_{\rm 3d}^\mu = \frac{1}{2\pi}\epsilon^{\mu\nu\rho}\partial_\nu a_\rho\label{3dcurrent}\ee
where the 3d fermion $\psi_{\rm 3d}$ is related to the 2d fermion $\psi$ in \eqn{freefermion} by the rescaling $\psi = \sqrt{2\pi R}\,\psi_{\rm 3d}$.
Upon reduction on the circle, at low-energies this descends to a map between currents of 2d theories, 
\be J_V^i =\bar{\psi}\gamma^i \psi\ \ \longleftrightarrow\ \ J^i_V = \frac{1}{2\pi}\epsilon^{ij}\partial_j\theta\label{2dcurrent}\ee
This, of course, is the usual bosonization map in two dimensions, with the Wilson line \eqn{wline} playing the role of the dual boson. We will also need one further fact: the 2d fermionic theory enjoys an enhanced symmetry in the IR when $m_F=0$. This is the axial symmetry, under which the fermion zero mode is invariant, but the tower of KK modes are not. But this too is easily mapped onto the bosonic theory,
\be J_A^i = \bar{\psi}\gamma^i\gamma^3\psi = -\epsilon^{ij} J_V^j
\ \ \longleftrightarrow\ \  J_A^i = \frac{1}{2\pi}\partial^i\theta
\label{axial}\ee
a result which is again familiar from 2d bosonization.  This is the current that arises from a shift symmetry of $\theta$. In the present context, it tells us that when $m_B^2$ is tuned to a critical value, which we take to be $m_B^2=0$, then no potential of the form \eqn{kkpot} can be generated for the Wilson line, despite our weak coupling intuition. Furthermore the fact that the Wilson line acts as a dynamical theta angle -- which would give it a mass of order $e_{\rm 2d}$, even in the absence of a KK-generated potential -- is similarly misleading. Instead, when $m_B^2=0$, the mere existence of the current \eqn{axial} is sufficient to tell us that the Wilson line remains gapless. This is the dual boson. 

\para
The argument above, which relies only on symmetries and the assumption of the duality, ensures that, at low-energies, the dynamics of the strongly coupled 2d theory takes the form
\be S_{\rm 2d} = \int \rmd^2x\ \frac{\beta^2}{2} (\partial_i\theta)^2\label{2dwilson}\ee
The question is: how can we fix $\beta^2$? We know from standard bosonization story, recounted in the previous section, that the free fermion point corresponds to $\beta^2  = 1/4\pi$.

\para
In fact, this too is fixed by the symmetries and, in particular, the normalisation of the axial current \eqn{axial}. To see this, suppose that we couple the fermion to a background vector field $A_i$. Upon compactification, this couples to the Wilson line as
\be S_{\rm 2d} = \int \rmd^2x\ \frac{\beta^2}{2} (\partial_i\theta)^2  + \frac{\theta}{2\pi}F_{01}\nn\ee
The equation of motion for $\theta$ is
\be \partial^2\theta = \frac{1}{2\pi\beta^2}F_{01}\ \ \ \Rightarrow\ \ \ \partial_i j_A^i  = \frac{1}{4\pi^2 \beta^2} F_{01}\nn\ee
which agrees with the anomaly for a free fermion, $\partial_i j^i_A = 2\times F_{01}/2\pi$, only when $\beta^2 = 1/4\pi$ as required.\footnote{An aside: the phrasing above makes it appear as if the anomaly coefficient depends on the marginal parameter $\beta^2$. This is misleading; instead, the normalisation of the axial current depends on the marginal parameter $\beta^2$. In particular, the normalisation of the fermionic axial current differs from \eqn{axial} in the presence of Thirring-like interactions.}


\para
As shown by Coleman \cite{coleman}, the bosonization map \eqn{2dcurrent} between currents is sufficient to understand how to deform the theory away from the free fermion fixed point, with interactions induced on both sides by $(j_V)_i(j_V)^i$ couplings.
We can also  turn on mass deformations in the theory.   The duality map tells us that turning on a Dirac mass $m_F$ for the fermion is equivalent to a mass $m_B^2 \sim -m_F$. (This equation is understood to hold at the fixed point, rather than in the UV theory.) The Dirac mass breaks the axial symmetry, which no longer prohibits the potential for $\theta$. The lowest dimension operator consistent with the remaining symmetries is $\cos\theta$. This is simply the usual 2d bosonization story.

\para
We could also consider adding a Majorana mass. In 3d, this is associated to the fermion bilinear $\psi_{\rm 3d}\psi_{\rm 3d}$ which, under bosonization, maps to
\be \psi_{\rm 3d}\psi_{\rm 3d}\ \ \  \longleftrightarrow \ \ \ {(\phi \cal M)}^2 \nn\ee
where ${\cal M}$ is the monopole operator in the bosonic 3d theory, which is dressed with $\phi$ to ensure gauge invariance in the presence of the Chern-Simons term. Compactifying and matching charges, the monopole operator descends to the vortex operator $e^{i\tilde{\theta}}$ where $\tilde{\theta}$ is the dual of the Wilson line: $\partial_i\theta = 2 \epsilon_{ij}\partial^j\tilde{\theta}$. The 2d Majorana mass is then
\be \psi\psi \ \ \ \longleftrightarrow\ \ \ e^{2i\tilde{\theta}}\nn\ee
In the action \eqn{2dwilson}, in which the compact boson has radius $\beta$, the dimension of the Dirac mass term $e^{i\theta}$ is  $1/4\pi\beta^2$; that of the Majorana mass $e^{2i\tilde{\theta}}$ is $2^2\times  \pi \beta^2$. Hence $\beta^2=1/4\pi$ is the unique point at which both of these deformations have the same dimension, namely 1, as it should be for the free fermion.

\para
The map $\psi_{3d} \longleftrightarrow \phi{\cal M}$ raises  a puzzle upon reduction to two dimensions\footnote{We thank Max Zimet for raising this issue.}. In three dimensions, the fermion corresponds to a monopole operator dressed with $\phi$, due to the Chern-Simons term. Yet in 2d our discussion above suggests that the fermion is associated purely with the Wilson line. In fact, this need not be the case. Consider the 2d theory which 
comes from a reduction of the 3d kinetic terms (ignoring potential terms, for which the classical analysis described here would appear to be woefully misleading). We start with the 2d XY model, with periodic scalar $\sigma\in [0,2\pi)$ (this is roughly the phase of $\phi$) whose shift symmetry is gauged.  The gauge field is subsequently  coupled to a dynamical axion $\theta$, a remnant of the 3d Chern-Simons term:
\be  S_{\rm 2d} = \int \rmd^2x \ \frac{\tilde{\beta}^2}{2}(\partial \sigma - a)^2 + \frac{1}{2\pi} \theta f_{01}  \nn\ee
This theory shares features with its 3d cousin, but sits in the same universality  class as the gapless boson. To see this, we dualise the periodic scalar
 $\partial_i\sigma=  \epsilon_{ij}\partial^j \tilde{\sigma}/2\pi\tilde{\beta}^2$, and write the action as
\be S_{\rm 2d} = \int \rmd^2x\ \frac{1}{8\pi^2\tilde{\beta}^2} (\partial\tilde{\sigma})^2 + \frac{1}{2\pi} (\theta- \tilde{\sigma}) f_{01}\nn\ee
The equation of motion for the gauge field then imposes $\tilde{\sigma}=\theta$. This shows that the XY sector is intimately connected to Wilson line, in analogy with the dressing of the monopole operator in 3d.

\para
Finally, we discuss the parity symmetries of our theory. The action of parity in 2d is $P_{\rm 2d}: \theta \mapsto -\theta$, so the Dirac mass term is invariant. A slightly more subtle question concerns the action of parity in 3d, which we take to be defined by $P_{\rm 3d}: x^2 \mapsto -x^2$. In the bosonic theory, this is a quantum symmetry that is not manifest at the level of the classical action, so in order to study it we turn to the dual theory. Upon compactifying the fermionic side, one finds that $P_{\rm 3d}$ descends to ${\bf Z}_2\subset U(1)_A$, under which the Dirac mass term is odd. The bosonization map then tells us that 3d quantum parity acts as $P_{\rm 3d}: \theta \mapsto \theta + \pi$. In other words, the Wilson line $e^{i\theta}$ is odd under $P_{\rm 3d}$. 

\para
Our determination of the bosonic theory \eqn{2dwilson} relied only on the identification of the symmetries, rather than any explicit 3d dynamics.  One might ask whether its possible to determine the value of $\beta^2$ directly from the 3d theory, without recourse to the duality. We have not been able to do this; indeed, successfully determining $\beta^2$ from first principles would present   strong evidence for the 3d bosonization duality. 

\subsubsection*{Addendum: A ${\bf Z}_2$ Gauge Field}

After this paper was accepted for publication, we realised that the discussion above is incomplete. In 2d, a free Dirac fermion is not equivalent to a compact boson. Instead, it is equivalent to a compact boson coupled to a ${\bf Z}_2$ gauge field. This ${\bf Z}_2$ gauge field should emerge from the reduction of the 3d $U(1)_1$ Chern-Simons theory\footnote{We're grateful to Shauna Kravec and John McGreevy for pointing this out to us.}.

\para
Indeed, it does. Key to understanding this is the subtle way in which  $U(1)_1$ Chern-Simons theory depends on the background spin structure, and how this is captured by a 2d topological term known as the Arf invariant. These issues are discussed in more detail in Appendix C of the subsequent paper \cite{arf}.

%
%
%
%

\section*{Acknowledgements}

We thank Nick Dorey, Sergei Gukov, Kristen Jensen, Shamit Kachru, Shiraz Minwalla, Mehrdad Mirbabayi and Max Zimet for a number of very useful discussions. 
We are supported  by the U.S.~Department of Energy under Grant No.~DE-SC0011637
 and by the STFC consolidated grant ST/P000681/1. DT is a Wolfson Royal Society Research Merit
Award holder.  CPT is supported by a Junior Research Fellowship at Gonville \& Caius College, Cambridge.

\end{document}